\documentclass[%
 reprint,
 amsmath,amssymb,
 aps,
]{revtex4-1}
\usepackage[english]{babel}
\usepackage[dvips]{graphicx}
\usepackage{dcolumn}
\usepackage{bm}

\begin{document}

\preprint{APS/123-QED}

\title{Melting of Rare-Gas Crystals: Monte Carlo Simulation versus Experiments}

\author{V. Bocchetti}
 \email{virgile.bocchetti@u-cergy.fr}
 \author{H. T. Diep}%
 \email{diep@u-cergy.fr, corresponding author}
\affiliation{Laboratoire de Physique Th\'eorique et Mod\'elisation, \\Universit\'e de Cergy-Pontoise, CNRS, UMR 8089\\
2, Avenue Adolphe Chauvin, 95302 Cergy-Pontoise Cedex, France}






\date{\today}

\begin{abstract}
We study the melting transition in crystals of rare gas Ar, Xe, and Kr by the use of extensive
Monte Carlo simulations with the Lennard-Jones potential. The  parameters of this potential have been deduced by Bernardes in 1958 from experiments of rare gas in the gaseous phase.  It is amazing that the parameters of such a popular potential were not fully tested so far.   In order to carry out precise tests, we have written a high-performance Monte Carlo program which allows in particular to take into account correctly the  periodic boundary conditions to reduce surface effects and to reduce CPU time. Using the Bernardes parameters, we find that the melting temperature of several rare gas is from 13 to 20\%  higher than that obtained from experiments. We have throughout studied the case of Ar by examining both  finite-size  and  cutoff-distance effects.  In order to get a good agreement with the experimental melting temperature, we propose a modification of these parameters to describe better the melting of rare-gas crystals.
\begin{description}
\item[PACS numbers: 34.20.Cf, 64.70.Dv, 65.20.+w, 65.40.-b]
\end{description}
\end{abstract}

\maketitle


\section{\label{sec:level1}INTRODUCTION}
Melting of crystals has always been a fascinating subject for more than a century since the discovery of the
empirical Lindemann's criterion \cite{Linderman1910}. The Lindemann's
criterion says that if the average of vibration amplitude
u, namely
$\sqrt{\langle u^2 \rangle}$ , exceeds a certain value, usually 10$\%$ of
the distance between nearest-neighbors, then the melting occurs. Times and over again, many authors have
tried to find out microscopic precursor mechanisms that
lead a crystal to melt. Until 30 years ago, one of the
favorite pictures of melting is the softening of a phonon
mode due to the temperature (T ). The atoms have no
longer restoring forces which keep them staying close to
their equilibrium positions: they move around and the
system goes to a liquid state. The soft-mode picture has
encountered some scepticism because in real crystals as
well as in simulations one observes that well below the
melting temperature ($T_m$ ), many defects, dislocations,
interstitial atoms ... are excited. Therefore, it is hard
to believe that the system stays in a periodic structure
with propagating phonon modes up to $T_m$ . Evidence of
defects is found in many works \cite{Nelson1979,Gomez2001,Gomez2003,Mermin1968}.
Another question that is unsolved in a clear manner is
the form of the potential that binds the atoms together
in a given lattice structure. In a microscopic point of
view, the potential should come mainly from the symmetry of atomic orbitals. But ab-initio calculations are
still far away from being able to use realistic hypotheses \cite{Pettifor1995}.
Empirical potentials have been used instead to
study melting. One can mention the popular 6-12 power
Lennard-Jones (LJ) potential \cite{Kittel2008,Aschrof1976}, various similar
power potentials, the many-body Gupta's potential \cite{Gupta1981},
the Stillinger-Weber (SW) potential  \cite{Stillinger1985}, and the Tersoff
potential \cite{Letters1986,Review1989}. Two-body potentials such as the LJ one
crystallize atoms in the FCC at low temperatures and
nothing else; this comes from the fact that LJ potential
is isotropic so the atoms are crystallized in the most dense isotropic structure, namely
the FCC lattice. In order to stabilize other structures, several
phenomenological potentials have been introduced, often without
a microscopic justification. For example, the  SW potential or the Tersoff potential
stabilize the diamond structure at low temperatures.
These potentials have been used with success to calculate
properties of Si clusters \cite{Dinda1995} and amorphous Si crystals \cite{Vink2001}.

In this paper, we use the LJ potential to study the
melting of rare-gas crystals which have the FCC lattice
structure at low temperatures. It is amazing that such a
simple question was not studied with precision so far in
spite of an abundance of experimental data on rare gas
such as Ar, Xe, Ne and Kr. Most of the melting studies
concerning rare gas were done in particular cases: small
clusters \cite{Vink2001}, adlayers on a substrate, etc. The main
reason to avoid to study the bulk melting may be due
to some technical difficulties such as periodic boundary
conditions, volume expansion with temperature, etc. Previous Monte Carlo (MC) studies of bulk melting
have been carried out with LJ potential but emphasize was put on the melting mechanism
rather than on the precise melting temperature in real materials \cite{Gomez2001,Gomez2003}.

The purpose of this paper is therefore to test whether or not the experimental $T_m$ can be reproduced by MC simulation using the values of the LJ parameters deduced for rare gas in the gaseous state long time ago\cite{Bernardes1958}.  We will show here that by appropriate choices of technical procedures, we are able to obtain melting temperature for various rare gas {\it directly} from our simulations, unlike previous simulations \cite{Morris,Errington,Matsny2005,McNeil-Watson} which have had recourse to various means and some thermodynamic functions to deduce it.  We find in this work the melting temperatures for several rare gas higher than experimental values.  A revision of the values of LJ parameters widely used in the literature for more than 50 years should be made in order to better describe the solid state of rare gas.  Note that in a recent work \cite{Matsny}, a  hypothetical thermodynamic integration path is used to find the relative free energies of the solid and liquid phases, for various system sizes, at constant cutoff radius, in an attempt to explain the overestimate of the melting temperature with the LJ potential.  However, due to various approximations, several results were not physically clear, in particular why the melting temperature oscillates with increasing cutoff distance.

Section \ref{sec:level2} is devoted to a description of the model and our MC technique. The results are shown and discussed in Sect. \ref{sec:level3}. Concluding remarks are given in Sect. \ref{sec:level4}.

\section{\label{sec:level2}MODEL AND METHOD}
The LJ potential is given by
\begin{equation}
U = \frac{1}{2} {\sum_ {{i \ne j}}{V_{ij}}}
\end{equation}
with
\begin{equation}
V_{ij} = 4\epsilon \left[ a\left(\frac{\sigma}{r_{ij}} \right)^{12}-b\left(\frac{\sigma}{r_{ij}} \right)^{6} \right]
\end{equation}
where $\epsilon$ , $a$, $b$ and $\sigma$ are constants, $r_{ij}$  denotes the
distance between two atoms at $\textbf{r}_i$ and $\textbf{r}_j$ . Here $a=1$ and $b=1$.
We list for convenience the values of the above constants for
various rare gas and their melting temperatures, using the data from Ref. \cite{Kittel2008}. Note that the
constants $\epsilon$ and $\sigma$ are those deduced with some
approximations using experimental data of the gaseous phase of
rare gas which have been described in the pioneer paper of Bernardes\cite{Bernardes1958}.
The listed values of
the constants should be therefore viewed as approximate values.
To our knowledge, there were not papers using these constants  to
verify if the melting temperature is correctly obtained, either by theoretical calculations or by simulations. The absence of works dealing with this point has motivated our present work.

In one of our previous works \cite{Gomez2003}, we have used
the LJ potential to investigate the mechanism which initiates the melting. We have counted the number of defects
created with increasing temperature ($T$) and found that the melting occurs when these defects interact with each other forming
a kind of chains of defects which break the solid periodic
state at a temperature well below the melting point.  In that work, we did not use the constants of the LJ potential for any specific material. We used instead the reduced temperature and dimensionless parameters. It was not our purpose to clarify the value of melting temperature for a given kind of crystal. It is now the time to verify those constants to see whether or not they yield the correct value of $T_m$ for the melting of the solid phase of rare gas.

\begin{figure}[!h]
\includegraphics[width=8cm,height=5cm]{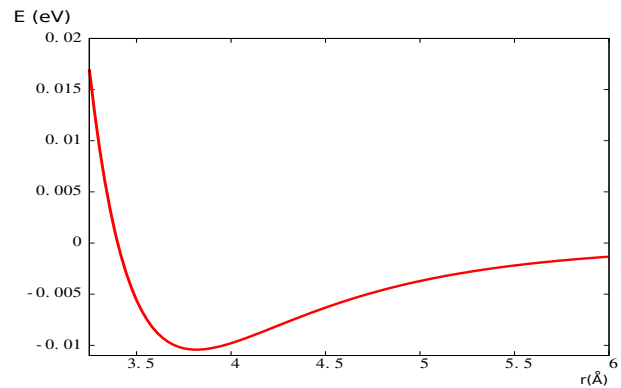}
\caption{\label{system}(Color online) The Lennard-Jones potential for Ar with the parameters listed in Table \ref{tab:table1}.}
\end{figure}

\begin{table}[h]
\caption{\label{tab:table1}%
Lenard-Jones parameters \footnote{Reference \cite{Kittel2008}}}
\begin{ruledtabular}
\begin{tabular}{llll}
\textrm{Element}&
\textrm{$\epsilon$ (eV)}&
\textrm{$\sigma$(\AA)}&
\textrm{$T_m\ $(K)}\\
\colrule
Ar&0.01042332126 & 3.40&84\\
Kr&0.01404339691 & 3.65&117\\
Ne&0.00312075487& 4.74&24\\
Xe&0.01997283116& 3.98&161\\
He&0.00087381136& 2.56&-\\
\end{tabular}
\end{ruledtabular}
\end{table}

Let us describe our MC technique in
the study of crystal melting. In our simulations we take a system of FCC lattice of size
$N=4\times L\times L\times L$ where $L=4$, 5, 6, 8 and 10 namely a system of $N=256\ $, 500, 1024, 2048 and 4000 atoms. Unlike spin systems on lattice where the system volume does not change with $T$,
a system of vibrating atoms expands its volume with increasing $T$. We use the method described in Ref. \cite{Allen1987} but
with a modification that the displacement of an atom $i$
from its current position ${\bf r}_i$ is taken in a sphere of very
small radius $r$ with two arbitrary angles ($\theta,\phi$). To respect the uniform spatial distribution around ${\bf r}_i$, we use
the Jacobian, namely we take an arbitrary $\cos(\theta)$ between
-1 and +1, instead of an arbitrary $\theta$. $\phi$ is taken at random between $0$ and $2\pi$. Note that we first move all atoms
to new positions, and change the system volume for a
small arbitrary amount, then calculate the new system
energy. We tune the magnitude of displacement ($r$) of the atoms and the volume variation in order to have an acceptation rate of displacement between $30\%\ $and $50\%\ $. This criterion is
empiric but it is frequently used in MC simulation. A MC step consists of moving all atoms, each with an arbitrary displacement,  and changing the system volume once. The transition probability to the new state is
 $\exp(-\frac{W}{k_B T}) $ where

\begin{eqnarray}
W = P\left(V_{new} - V_{old} \right) + \frac{1}{2} \left(U_{new} - U_{old}  \right) + \nonumber\\
N  k_{B}  T  \ln\left({\frac{V_{old}}{V_{new}}}\right)
\end{eqnarray}
where $P$ is the pressure which is set to zero here (constant
pressure), $V_{old}$ and $V_{new}$ are old and new system volumes, $U_{old}$ and $U_{new}$ old and new system energies.

One of the most difficult tasks in melting simulations is the application of the periodic boundary conditions. This is due to random positions of atoms at the crystal boundaries. We have taken the following actions to shorten, without loosing accuracy, our simulation CPU time:

(i) each atom has a list of neighbors up to a distance $r_d$ longer than the potential cutoff distance $r_c$. To establish the list for the first time, we have to calculate all distances and we select neighbors at $r\leq d$. It takes time.  The fact to choose $r_d>r_c$ is to ensure that for small displacements, some neighbors with $r_c<r<r_d$ can enter inside the cutoff sphere without the need to reestablish
the list of neighbors

(ii) when an atom moves outside a sphere of radius $r_1$ around its equilibrium position, its list of neighbors is updated by recalculating all pairs, neighbors  up to $r_d$ fill the new list. $r_1$ is chosen to be equal to 20\% of $r_0$ the nearest-neighbor distance at equilibrium ($T=0$)

(iii) periodic boundary conditions are applied by translating the system in all directions in a manner that avoids mismatches at the boundaries.

\begin{figure}[!h]
 \includegraphics[width=6cm]{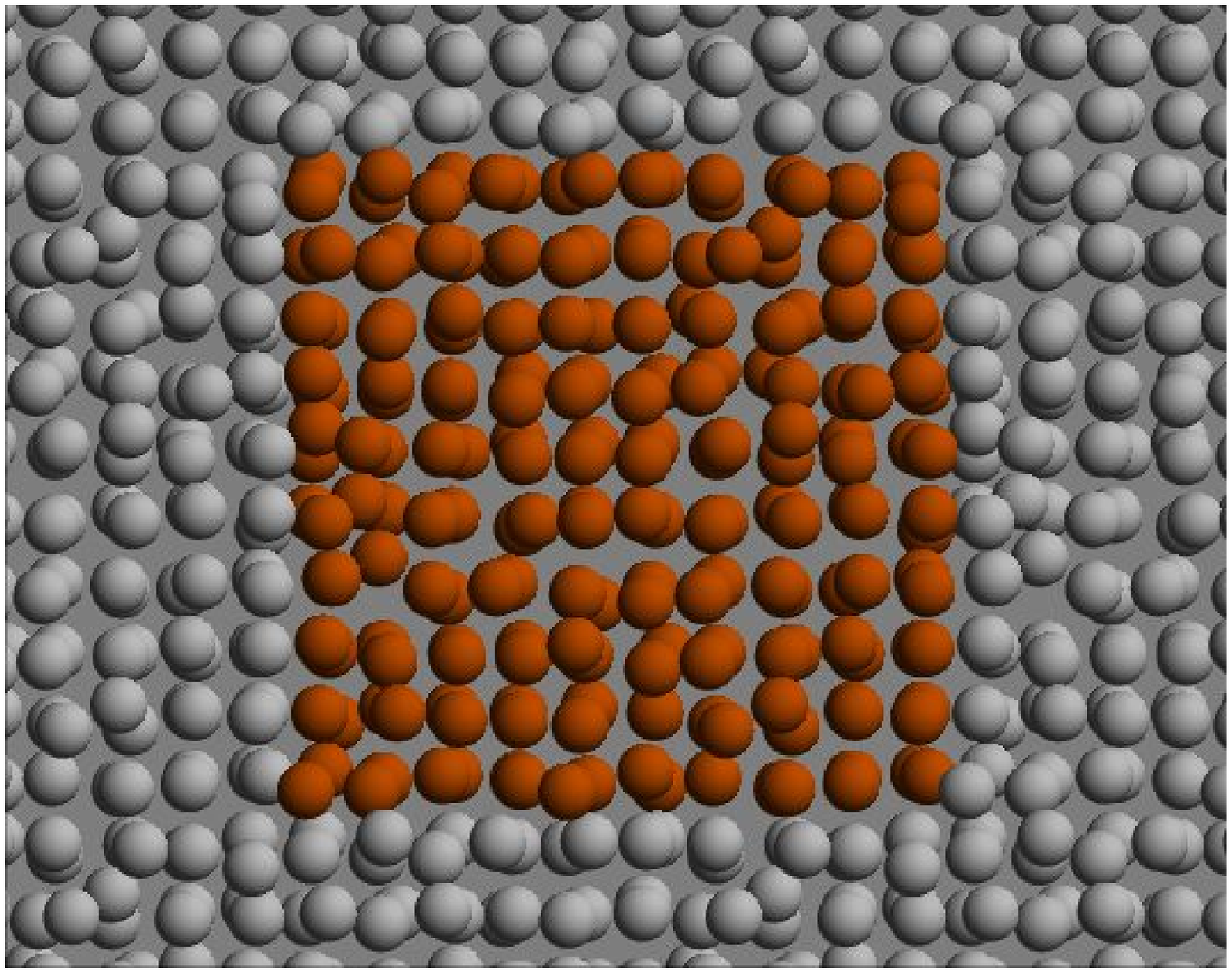}
  \includegraphics[width=6cm]{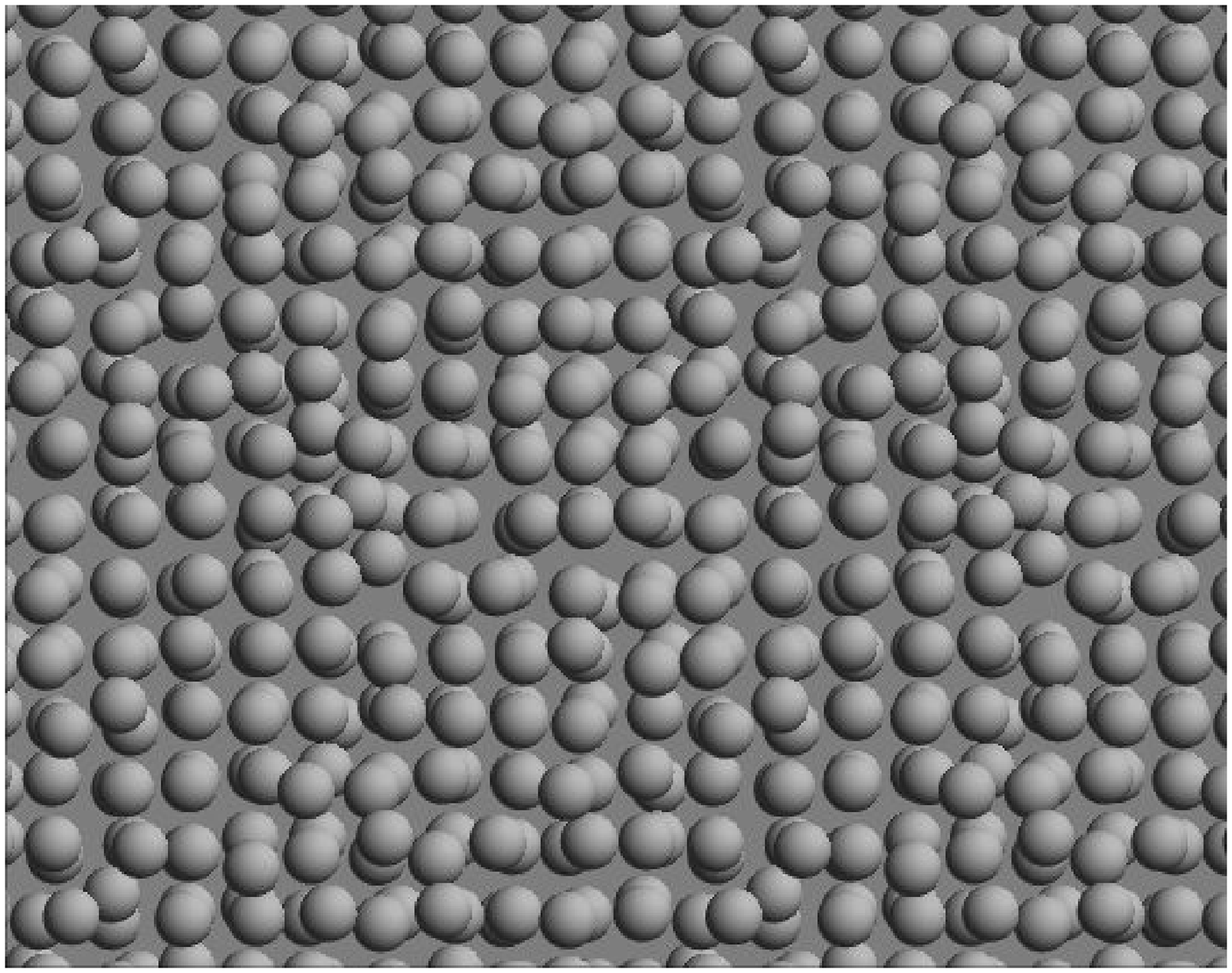}
\caption{\label{pbc}(Color online) Upper: Snapshot of the system (magenta) with its translated atoms (gray)  for periodic boundary conditions, Lower: The same snapshot of the system with its translated atoms  but with the same color (online): boundaries are undistinguishable.  Argon crystal at $T = 95$ K with $N=500$ atoms.}
\end{figure}

The last step is the most important technically. We show in Fig. \ref{pbc} a snapshot of the system and its translated atoms by periodic boundary conditions. We see that if they are of the same color, one cannot distinguish the system boundaries.

In the following, we will show results with a cutoff distance $r_c$ equal to twice $\ell$, the lattice constant of the FCC cell, namely  $r_c=2\ell=2r_0\sqrt{2}$ where $r_0$ is the nearest-neighbor distance.  As seen  in section \ref{cutoff} this value of $r_c$ is large enough to ensure a correct value of the melting temperature.

\section{\label{sec:level3}MELTING OF RARE-GAS CRYSTALS}
\subsection{The case of Ar}
\begin{figure}[!h]
 \includegraphics[width=8cm,height=5cm]{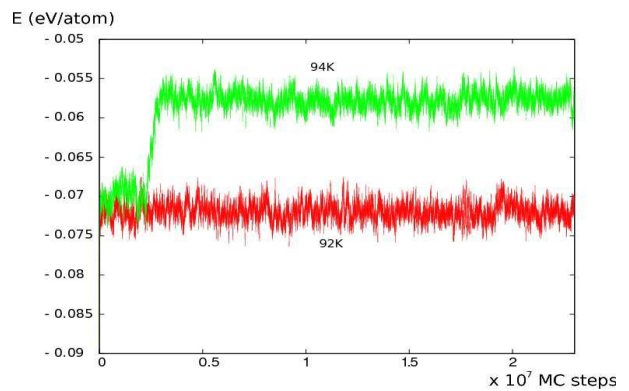}
\caption{\label{evtimekarg}(Color online) Energy per atom $E$ versus number of MC steps per atom $t$ for an Argon crystal at $T = 92$ K (red) and $T = 94$ K (green) with $N=256$ atoms.}
\end{figure}

Let us show first the results for Ar obtained by using the values listed in Table \ref{tab:table1}.  We will discuss next the modification necessary for obtaining the result in agreement with experiment.
In order to take a correct average of physical quantities, we record spontaneous values of all physical quantities during each MC run. We have to go to several
millions of MC steps before observing statistical fluctuations around equilibrium. We show an example of the
energy per atom $E$ versus MC time in Fig. \ref{evtimekarg} for $N=256$ atoms at two temperatures $T=92$ K and 94 K.  At $T=92$ K, the system is still in the solid phase. Its energy is stabilized after about one million MC steps per atom. However, at $T=94$ K, the system melts after three millions of MC steps per atom: $E$ is stabilized in the liquid phase only after such a long MC time. It is very important to emphasize that the convergence time to equilibrium depends
on various MC parameters such as value of displacement
magnitude $r$, volume variation $\delta$ etc. So, nothing can replace an
observation of the time-dependence of physical parameters
during the simulation, although this consumes a huge computer memory.

\begin{figure}[!h]
 \includegraphics[width=8cm,height=5cm]{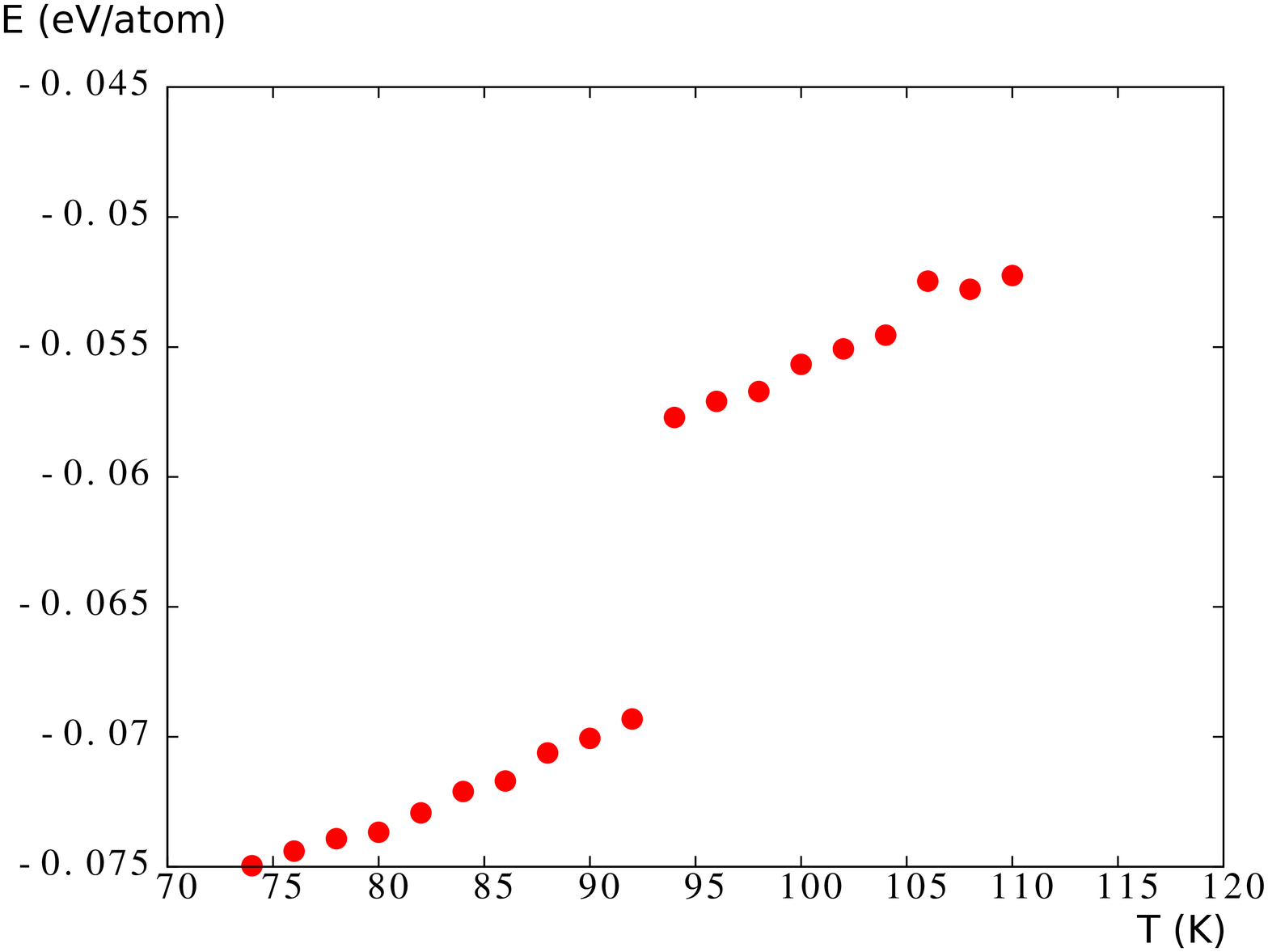}
\caption{\label{evtkarg}(Color online) Energy per atom $E$ versus temperature $T$ for an Argon crystal of 256 atoms, obtained
with  the parameters listed in Table \ref{tab:table1}. One observes that
 the melting temperature is $T_m \simeq 93$ K while the experimental value is $84$ K. Run of 50 millions MC steps/atom. See text for comments.}
\end{figure}

\begin{figure}[!h]
\includegraphics[width=2.8cm]{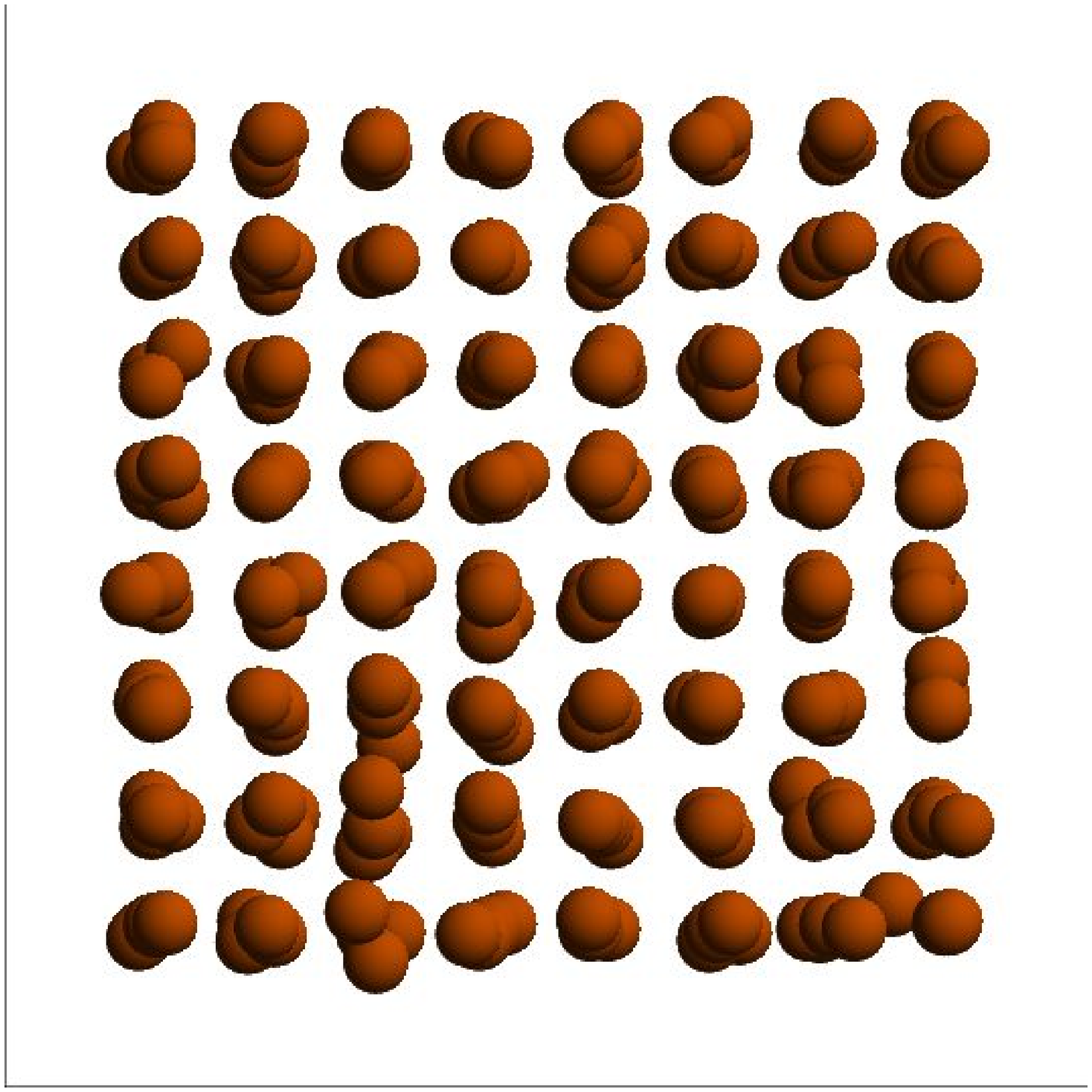}\hfill
\includegraphics[width=2.8cm]{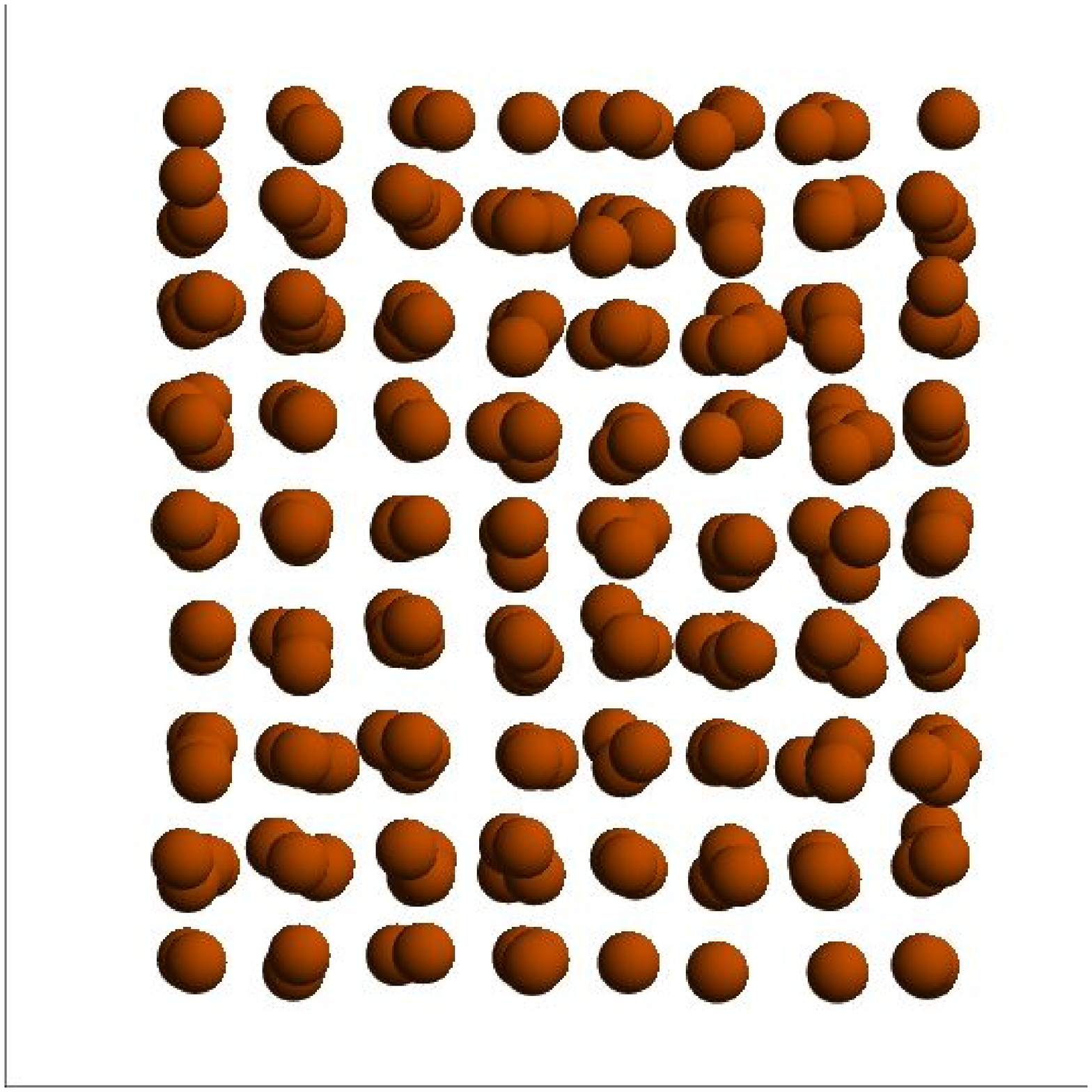}\hfill
\includegraphics[width=2.8cm]{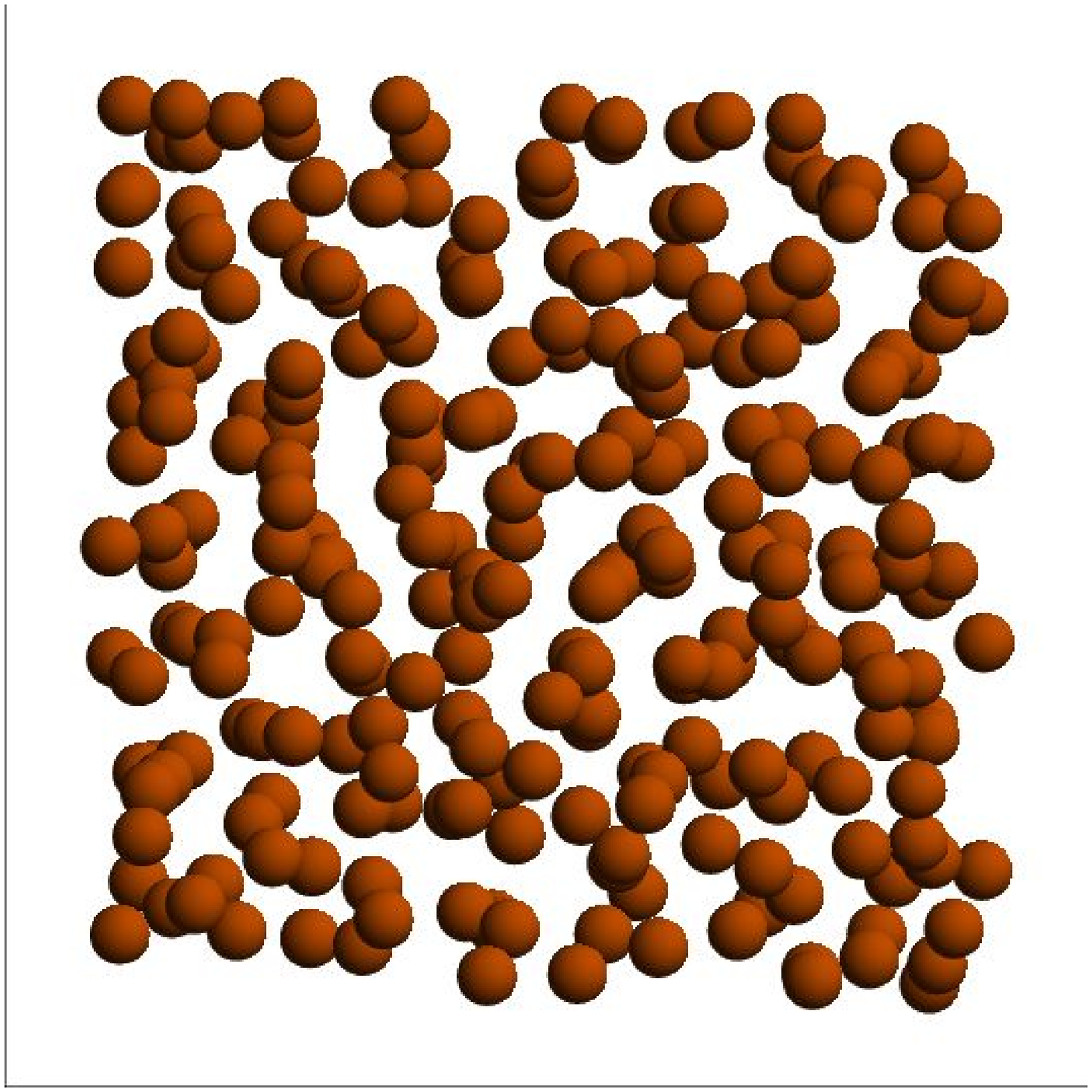}
\caption{(Color online)Instantaneous pictures of the supercell of Ar for different temperatures, respectively from left to right : 75 K (far below $T_m$), 92 K (close to $T_m$) and 100 K (above $T_m$).}\label{sys_pict}
\end{figure}

We show in Fig. \ref{evtkarg} the energy per atom $E$ versus $T$ in
the case of Ar with $N=256$ atoms, using the parameters given in Table \ref{tab:table1}. We
observe here that the melting occurs at $T_m \simeq 93$ K with a large latent heat. This value of $T_m$ is
rather far from the experimental data given in Table \ref{tab:table1}. We return to this point later. We show now the snapshots of the system for different temperatures in Fig. \ref{sys_pict} below and above the melting. As seen, the system just starts to be spatially disordered at 92 K, and is well in the disordered phase (liquid) at 100 K.

\begin{figure}[!h]
 \includegraphics[width=8cm,height=5cm]{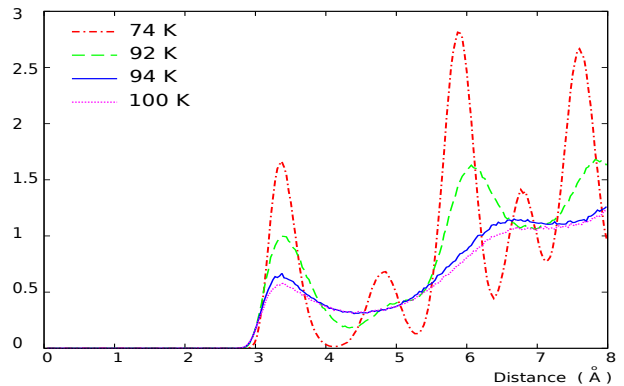}
\caption{\label{rdfarg}(Color online) Radial distribution of Ar for different temperatures: below melting $T=74$ K (discontinued red) and 92 K (discontinued green), above melting $T=94$ K (solid blue) and 100 K (solid magenta), with $N=256$ atoms.}
\end{figure}

The radial distribution for Ar is shown in Fig. \ref{rdfarg} at several
temperatures. One sees clearly the peaks at first, second
and third neighbor distances for $T < T_m$  indicating the crystalline phase, while there is
no such clear distinction for $T > T_m$ where a liquid phase
is set in.

\begin{figure}[!h]
 \includegraphics[width=8cm,height=5cm]{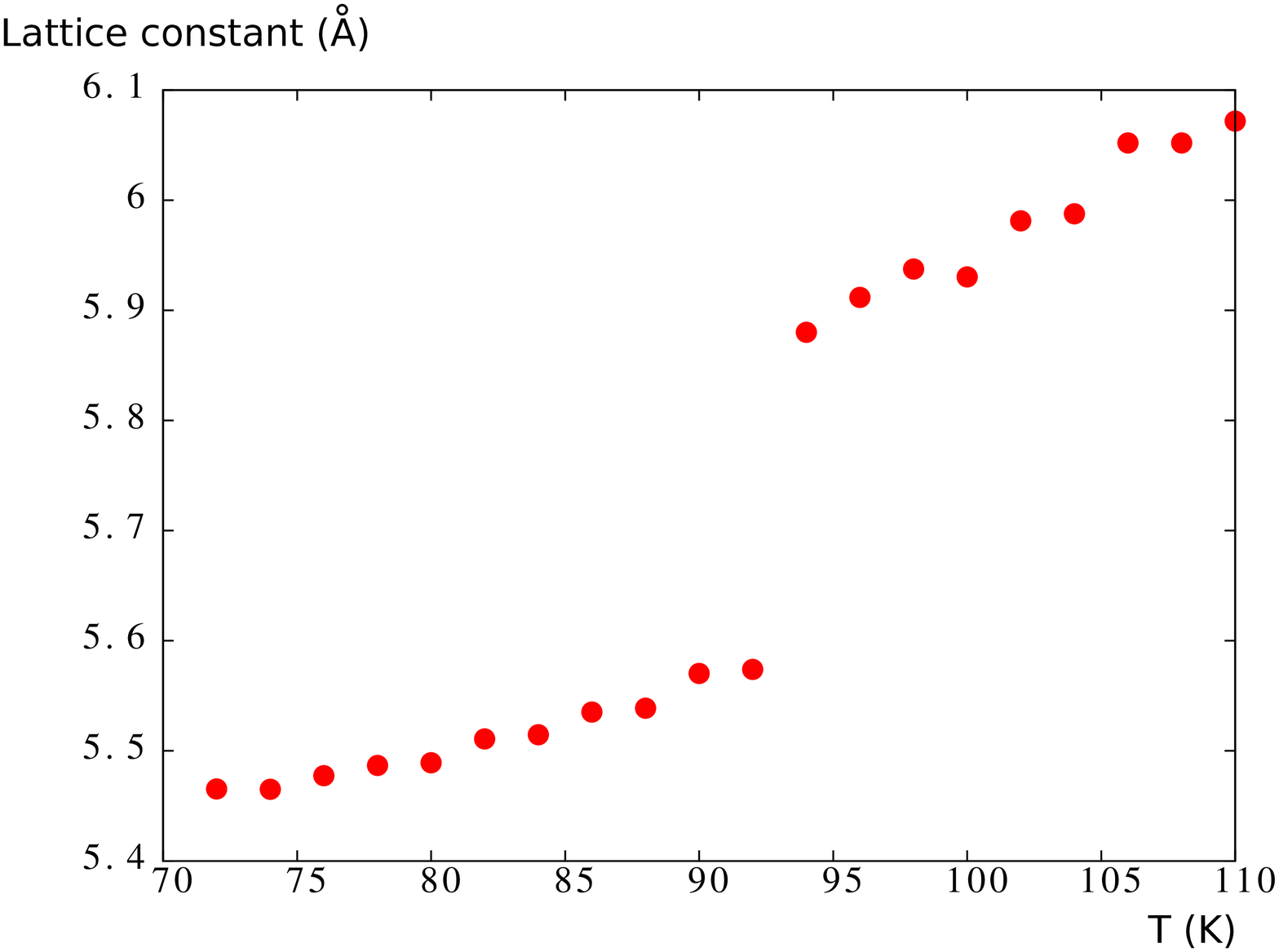}
\caption{\label{dilatation}(Color online)Evolution of the lattice constant against temperature $T$. As we can see, there is a jump of this quantity which occurs at the same temperature of the jump
of the energy in Fig. \ref{evtkarg}. Argon crystal, $N=256$.}
\end{figure}

We display in Fig. \ref{dilatation} the curve which shows the dilatation of the simulation box against the temperature.

Let us discuss now the difference between our value of $T_m(N=256)=93$ K with the experimental value $T_m$(exp)=84 K.   Clearly, the parameters given in Table \ref{tab:table1} which have been deduced with experimental data for gaseous Ar do not describe well the melting of solid Ar.   Note however that the above value of $T_m$ is for $N=256$ atoms.  If we consider larger samples, $T_m$ should increase further.

In order to modify correctly the LJ parameters while respecting the known properties of Ar,  we use the listed  values of LJ parameters to
estimate the size effect, namely to find the value of $T_m(N=\infty) $ for an infinite crystal.
Once this task is done, we then we look for the parameters ($\epsilon,\sigma$) which give correctly the experimental $T_m$(exp).  The effect of the cutoff distance will be examined.

\subsection{\label{finitesize}Size effect}
\begin{figure}[!h]
 \includegraphics[width=8cm,height=5cm]{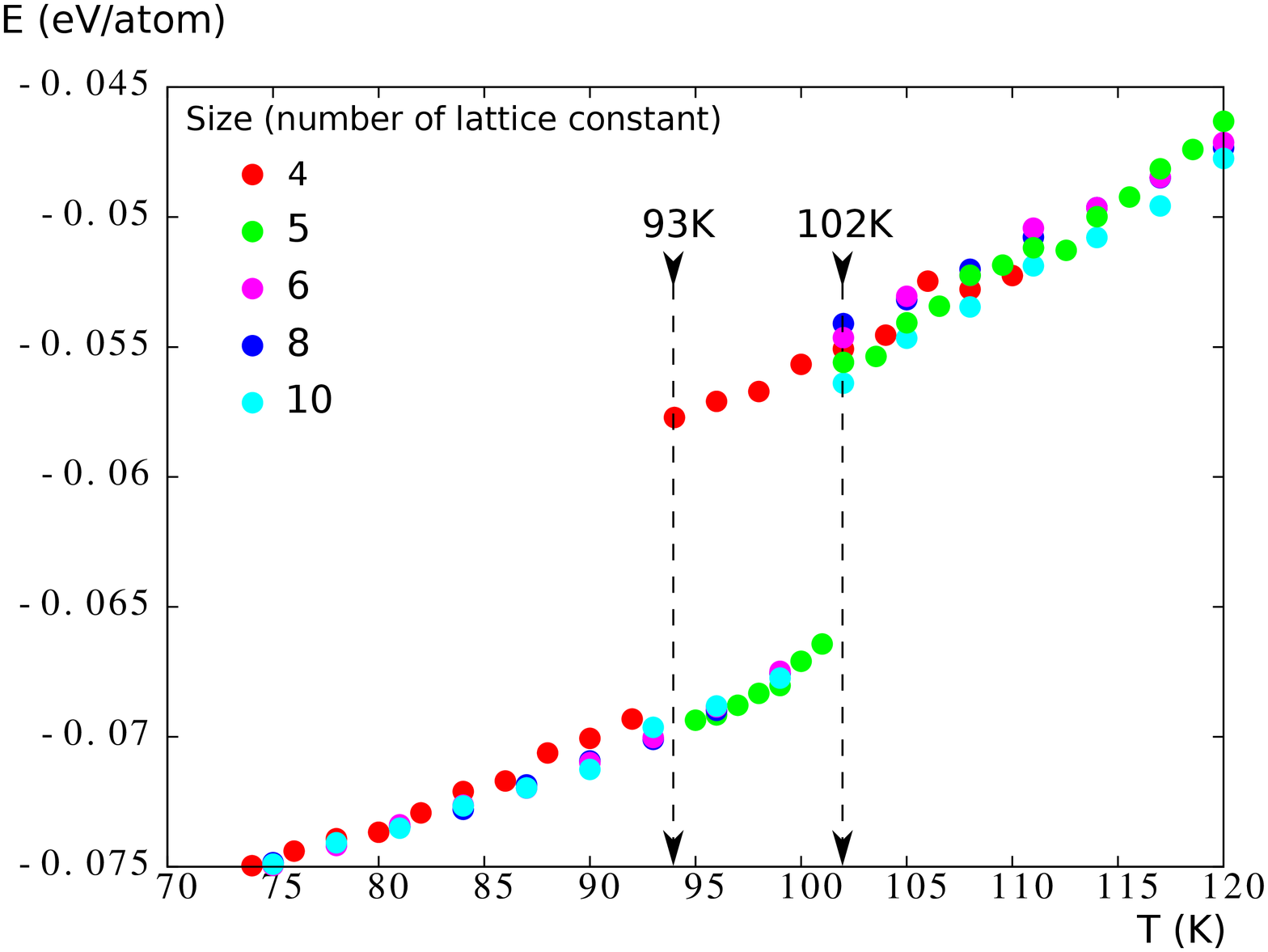}
\caption{\label{size_effect}(Color online)Energy per atom versus $T$ with various system sizes from $N=256$ (4 FCC lattice cells) to $N=4000$ (10 FCC lattice cells). The arrows indicate $T_m$ for the smallest and largest sizes. See text for comments. }
\end{figure}

The size effect is an important fact when we work with numerical simulations because of the finite size of the simulation cell. In second-order phase transitions, finite-size scaling laws allow us to calculate the critical exponents by varying the system size \cite{Binder,Hohenberg,Ferrenberg1,Ferrenberg2}.  The correlation length is infinite at the critical point in the thermodynamic limit.  However, in first-order transitions, the correlation length is finite at the transition point,  the two phases coexist\cite{Barber,Binder87} and the energy is discontinuous. The correlation is finite at the transition temperature means that when the system size exceeds the correlation length the transition temperature does not depend anymore on the system size.  We need not to go to the infinite limit to find saturated transition temperature.  Note that if the size of the system is too small, a first-order phase transition can appear as a second-order phase transition if the correlation length exceeds the system size.  As we can see in Fig.\ref{size_effect}, fortunately
when the size of the simulation cell is larger than $5$ lattice constants, melting temperature $T_m$
is bounded by $102$ K.  It means that the correlation length at the melting temperature is smaller than 5 lattice spacings.

The saturated melting temperature obtained by our MC simulations is thus $18$ K (or $20\%$) higher than the experimental one.
 This disagreement come from the values of the Bernardes parameters used for the LJ potential.  The reason is that
these values have not been calculated by fitting with
the melting temperature but they were calculated with a low-density gas using the second virial coefficient \cite{Bernardes1958}. The Bernardes parameters are therefore questionable as we can see in Refs. \cite{Klein1969,Batchelder1967,Bobbs1957}. Taken from those articles, the values of the nearest-neighbor distance, the cohesive energy and the bulk modulus calculated using the Bernardes parameters are listed in Table \ref{tab:table5}.  We can see the differences between experimental values and theoretical ones for all cases Ne, Ar, Kr and Xe. The Bernardes parameters also yield the high melting
temperature found in our simulations with respect to the experimental one.

\begin{table}[h]
\caption{\label{tab:table5}%
Experimental (Exp.) and theoretical (Th.) values of NN distance $r_0$, cohesive energy $u_0$ and bulk modulus $B_0$\footnote{Reference \cite{Klein1969,Batchelder1967,Bobbs1957}} for comparison. See text for comments.}
\begin{ruledtabular}
\begin{tabular}{llllll}
\colrule
&& Ne&Ar&Kr&Xe\\
\hline
$r_{0}$&Exp.& 3.13&3.75&3.99&4.33\\
&Th.& 2.99&3.71&3.98&4.34\\
\hline
$u_{0}$&Exp.& -0.02&-0.08&-0.11&-0.17\\
&Th.& -0.027&-0.0889&-0.12&-0.172\\
\hline
$B_{0}$&Exp.& 1.1&2.7&3.5&3.6\\
&Th.& 1.81&3.18&3.46&3.81\\
\end{tabular}
\end{ruledtabular}
\end{table}

At this stage, it is interesting to note that Molecular Dynamics (MD) simulation of melting of a perfect crystal with periodic boundary conditions produce superheating. There is an empirical rule which states that the melting temperature of a crystal without any surface and any defects, is $20 \% \ $higher than the true thermodynamic melting temperature $T_m$. However in MC simulations, defects and dislocations are naturally created in the crystal by means of random numbers used in every MC step for atom displacements. Thus, the superheating should not exist. Agrawal {\it et  al.} have shown  in Ref. \cite{Agrawal2005} that for Ar, with MC simulation, $T_m$ is about $15\%$ higher than the experimental value. With our results for Ar, we find an increase of about $20\%$.  For Kr, we find an increase of $13\% \ $ for $N=256$ atoms after $23\times 10^6$ MC steps per atom. This increase is more important if we consider larger sizes as seen in the case of Ar. As said, the high values of $T_m$ in MC simulations are not due to the superheating as in MD simulations. Rather, we believe that these high values are due to the inaccuracy of the listed Bernardes parameters. We will propose a modification in the following

\subsection{Modification of Lennard-Jones parameters}

In order to reduce the $T_m$ value, we propose now to modify the value of
$\epsilon$, the prefactor of the LJ potential, and the coefficient $\sigma$.

Note that in papers dealing with melting in other materials by means of MD calculations or MC simulations, there have been several propositions to modify constants appearing in potentials in order to obtain a correct agreement with experimental value of $T_m$. This is because these constants are often deduced from experimental data which are not valid for the whole temperature range.  Among these papers, we can mention the case of melting of Si crystal studied by MD \cite{Cook,Yoo} and MC simulations \cite{Agrawal2005,Bocchetti}, using the Tersoff potential \cite{Letters1986}.  Our proposition to modify some constants of the LJ potential when applied to a rare-gas crystal is certainly a necessity in order to reproduce the experimental $T_m$.

\begin{figure}[!h]
 \includegraphics[width=8cm,height=5cm]{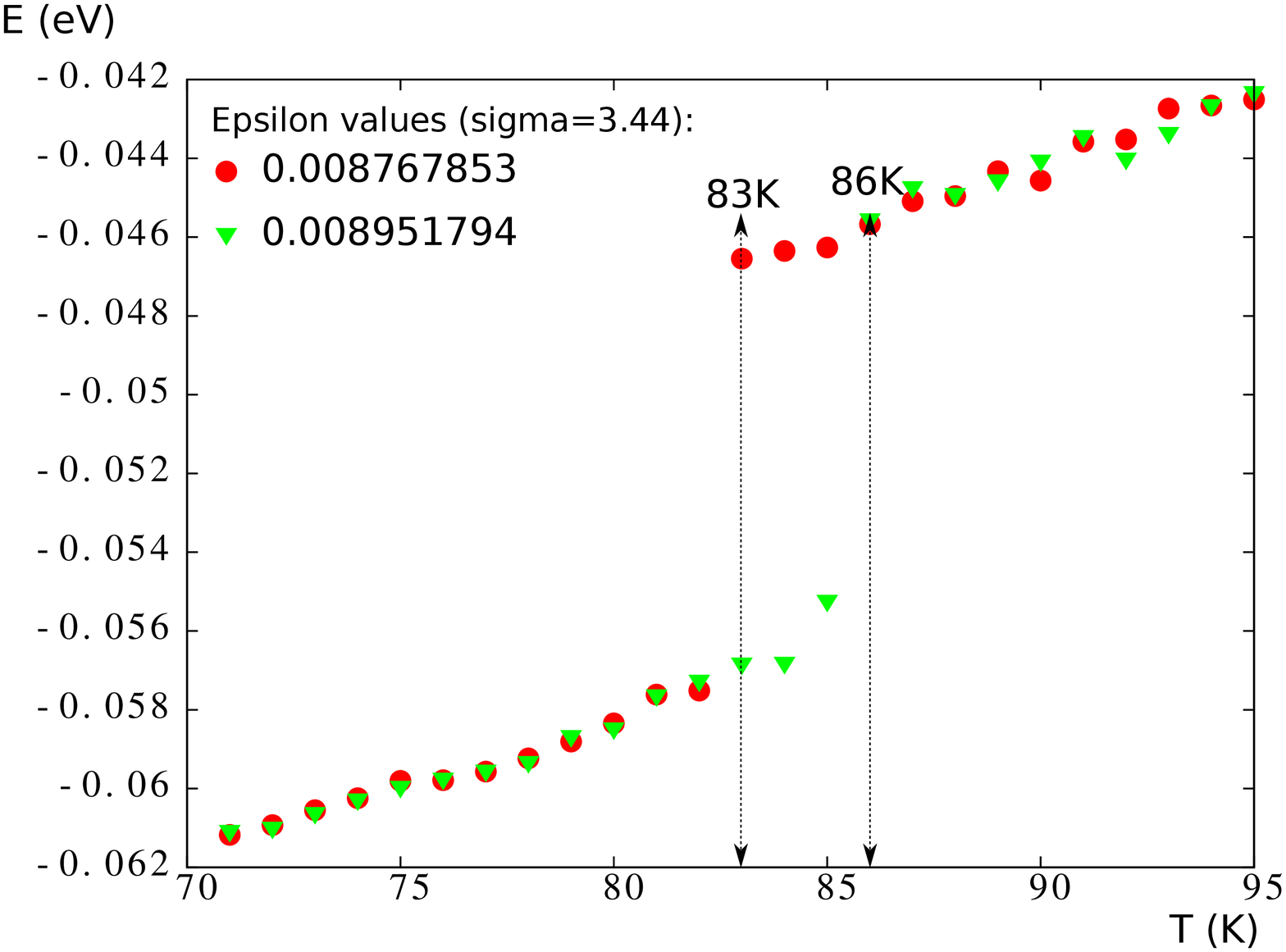}
\caption{\label{epsilon_effect}(Color online) Energy per atom versus $T$ for two selected values of $\epsilon$, with $\sigma=3.44$ \AA,  for an Ar crystal with $N=500$. The arrows indicate $T_m$ for the two indicated values of $\epsilon$. See text for comments. }
\end{figure}

We have done simulations with different pairs of $(\epsilon,\sigma$).  It turned out that $T_m$ depends essentially on $\epsilon$. There is however an optimal value of $\sigma$ which is $\sigma=3.44$ \AA \ corresponding to the experimental nearest-neighbor distance $r_0=3.75$ \AA \ of Ar (cf. Table \ref{tab:table5}).  We show in Fig. \ref{epsilon_effect} the curves obtained for two selected values of $\epsilon=$0.008767853 and
0.008951794 which give respectively $T_m=83$ K and 86 K.  These values of $T_m$ are in agreement with the experimental value 84 K within statistical errors.  Note that the modified $\epsilon$ is about 15\%  smaller than the original Bernardes value  $\epsilon=0.01042332126$.

\subsection{\label{cutoff}Effect of cutoff distance}

\begin{figure}[!h]
 \includegraphics[width=8cm,height=5cm]{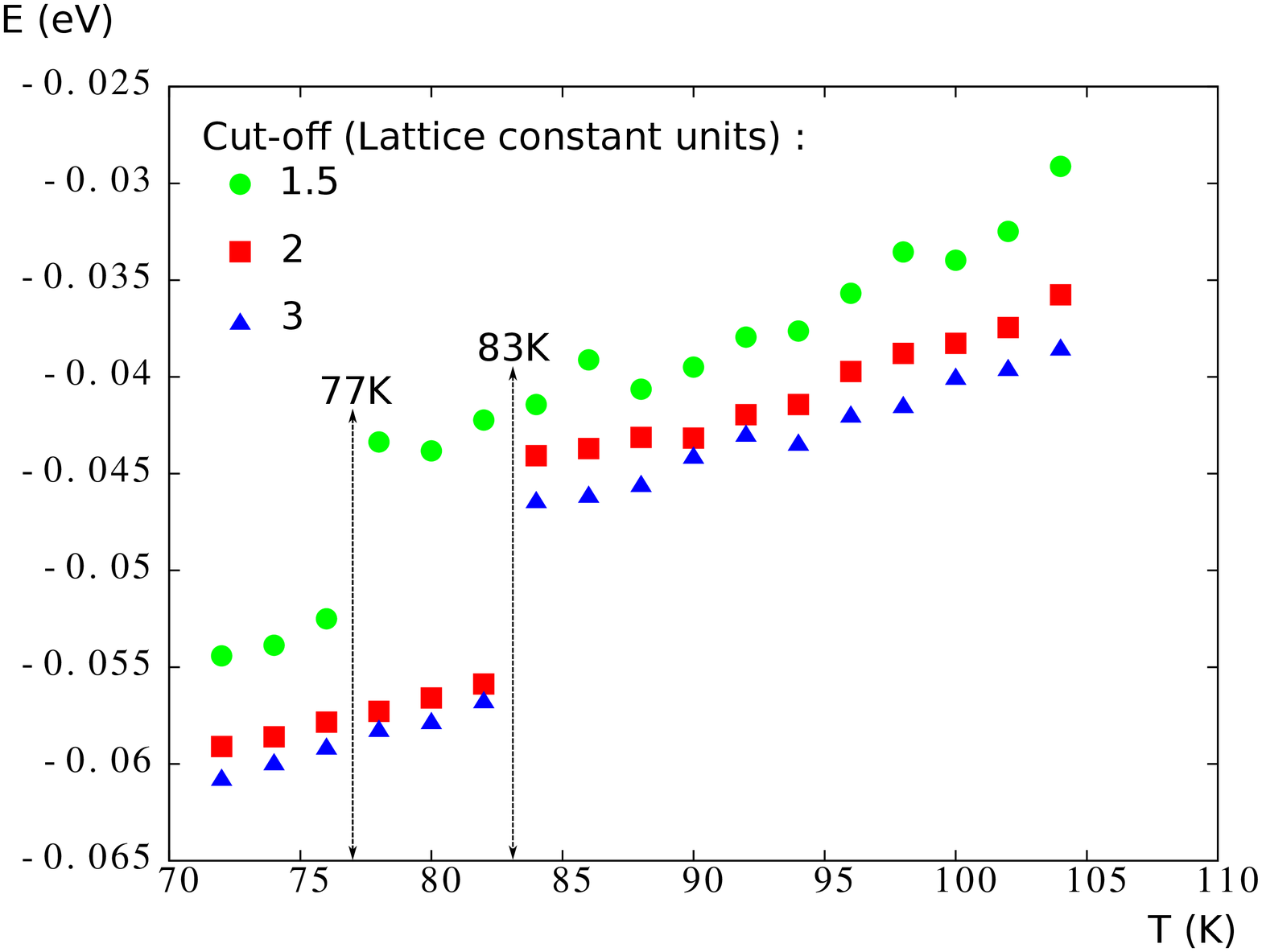}
\caption{\label{rc_effect}(Color online) Energy per atom versus $T$ with various values of $r_c$ for Ar crystal with $N=500$. The left arrow indicates $T_m$ for $r_c=1.5\ell$ and the right arrow indicates $T_m$ for $r_c=2\ell \sim 10.6$ \AA and $3\ell \sim 15.9$ \AA. Note that $\ell$ is the FCC lattice constant which is equal to $r_0\sqrt{2}$ where $r_0=3.75$ \AA \ is the NN distance.  See text for comments. }
\end{figure}

At this stage, a natural question we ask ourselves is "what is the effect of the cutoff distance $r_c$?". We know that for a long-range interaction, the longer the interaction range is the lower the energy becomes. As a consequence, the melting transition is higher.  However, as $r_c$ increases, the contribution of neglected neighbors becomes smaller.  From a certain value of $r_c$, $T_m$ does not vary significantly. This is observed in Fig.  \ref{rc_effect} where $T_m$ is saturated for $r_c\geq 2\ell$, i. e. $r_c \gtrsim 10.6$ \AA. All the results shown above for $r_c$ are valid in the discussion of the size effect and the modification of $\epsilon$ and $\sigma$.

\subsection{The case of other rare gases}
\begin{figure}[!h]
 \includegraphics[width=8cm,height=5cm]{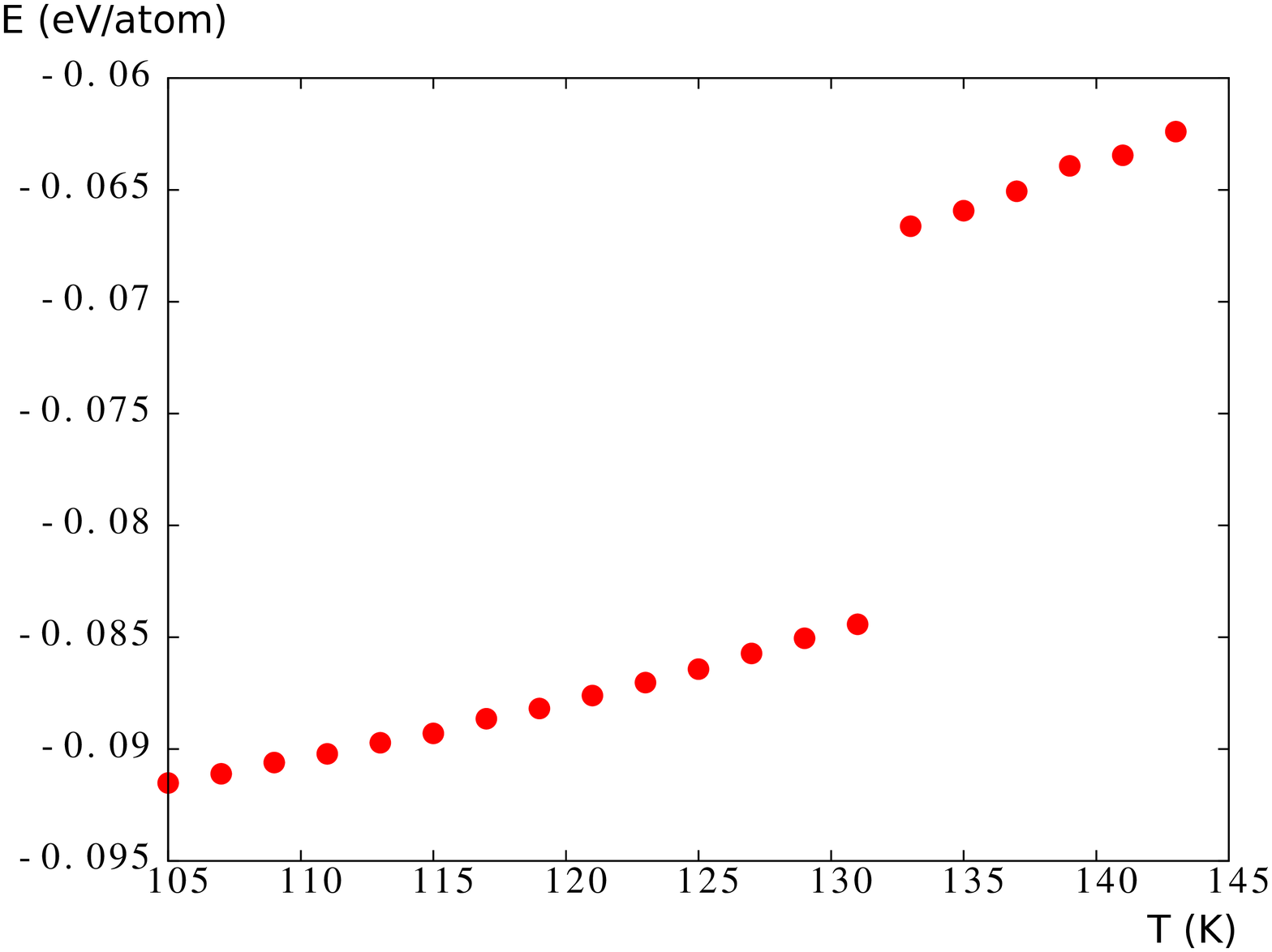}
 \includegraphics[width=8cm,height=5cm]{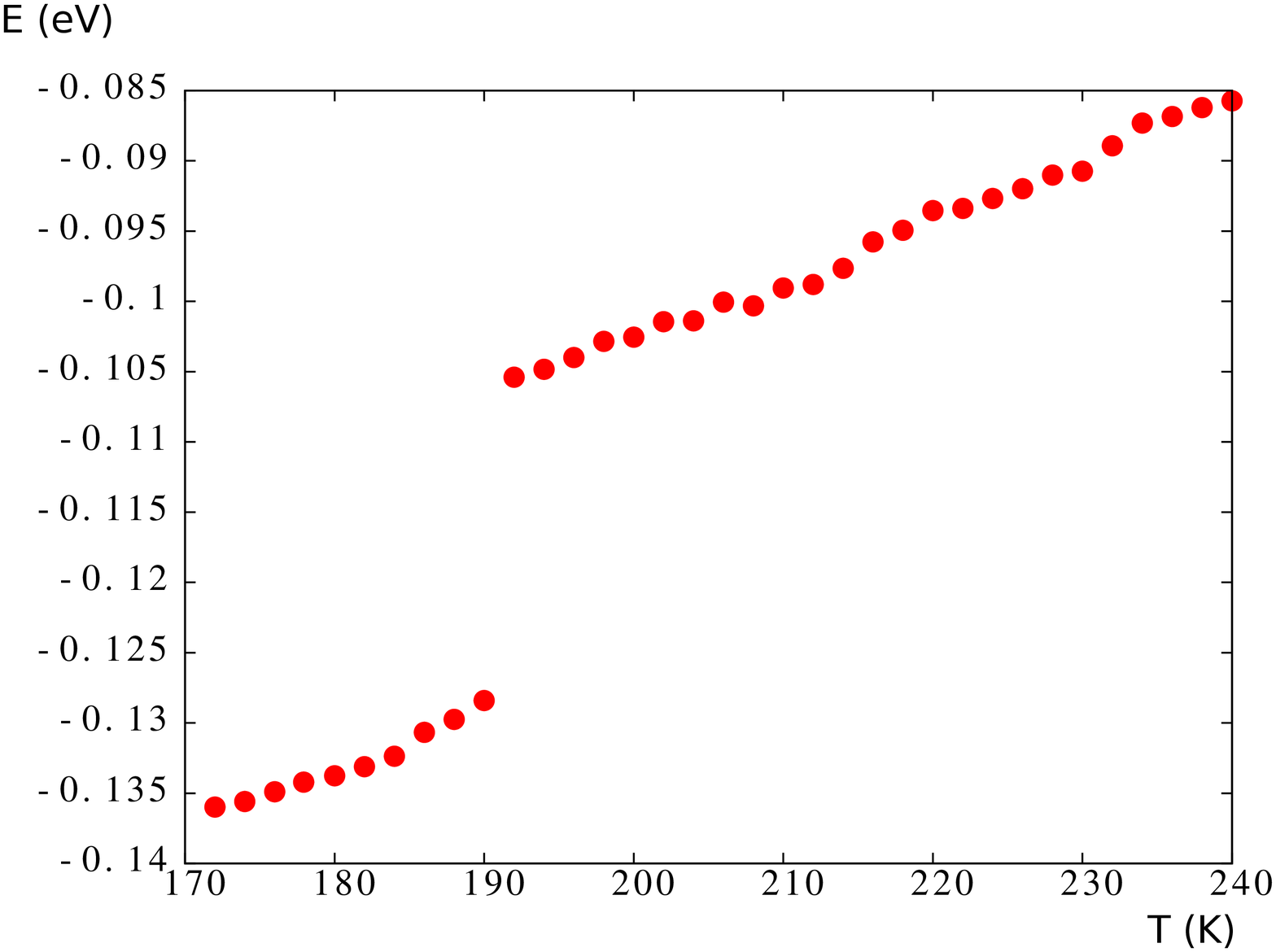}
\caption{\label{evtkr}(Color online) Upper: Energy versus temperature for a Krypton perfect crystal with $N=256$ atoms. One observes that $T_m = 132$ K while the experimental value is $117$ K. Lower: Energy versus temperature for a Xenon perfect crystal with $N=500$ atoms. One observes that $T_m = 191$ K while the experimental value is $161$ K. These curves have been obtained with the Bernardes values of parameters. }
\end{figure}

In order to show that our algorithm works well with other rare gas, we have plotted the curve of energy versus temperature, obtained for Krypton and Xenon in Fig. \ref{evtkr}.  Again here, we see that $T_m$, even for a small size, is already higher than the experimental value for each crystal.  We think that the Bernardes parameters for Kr and Xe should be modified to get an agreement with experiments as what proposed above for Ar.

\section{\label{sec:level4}CONCLUSION}

We have shown in this paper results of the melting temperature for rare gas, by performing extensive MC simulations  with the LJ potential.  We have obtained directly from our simulations physical quantities such as internal energy, lattice constant  and radial distribution as functions of temperature. We have shown that melting occurs with a large latent heat and a jump in the lattice constant. Effects of system size and cutoff distance have been investigated.  Let us emphasize that the theoretical LJ parameters widely used in the literature \cite{Bernardes1958} yield a melting temperature higher than
the experimental one as seen above, from $\sim$ 15\% for Kr to $\sim$ 20\% for Ar. This is not a surprise because those LJ parameters already yield theoretical NN distance, cohesive energy and bulk modulus different from corresponding experimental ones (see Table \ref{tab:table5}).
We have demonstrated that, in order to reduce the melting temperature to fit with experiments, it is necessary to modify the original Bernardes LJ parameter $\epsilon$  in such a way to reduce the energy at $T=0$. The effect of $\sigma$, within possible values of NN distance, is very small on $T_m$.   A good agreement on $T_m$ between experiments and simulations for Ar is obtained with the modified values given in Fig. \ref{epsilon_effect}. For other rare gas such as Kr and Xe, we have to proceed to a modification of their Barnardes parameters in a manner similar to that done above for Ar if we want to  get a good agreement with experiments.

{}

\end{document}